%% file: Document/arxiv_main.tex
\documentclass[letterpaper]{article}

\usepackage{fancyhdr}

\usepackage{xcolor}

\usepackage{graphicx}

\usepackage{gensymb}
\usepackage{amssymb}
\usepackage{amsmath}

\makeatletter 
\newif\if@anonymous 
\@anonymousfalse 

\usepackage[colorlinks=true, allcolors=blue]{hyperref}

\usepackage[left=2.58cm,right=2.58cm,top=2.58cm,bottom=2.58cm]{geometry}
\newcommand{\articletype}[1]{}

\renewcommand{\title}[1]{{\exhyphenpenalty=10000\hyphenpenalty=10000 
 \fontsize{18}{21}\selectfont\noindent\raggedright
        \textsf{#1}\par}\suppressfloats[t]}

\renewcommand{\author}[1]{{\vspace{5mm}%
   \fontsize{10}{12}
      \raggedright \if@anonymous Author list removed for anonymity \else #1 \fi
	  \vspace{3mm}}}

\newcommand{\affil}[1]{{\fontsize{8}{10}\selectfont
       \raggedright \if@anonymous \phantom{#1} \else #1 \fi}
	   }

\newcommand{\email}[1]{\vspace*{12pt}{\fontsize{8}{10}\selectfont
       \raggedright {\bfseries E-mail:} \if@anonymous \phantom{#1} \else #1 \fi}
	  \vspace{3mm} }
	   
\newcommand{\keywords}[1]{{\fontsize{8}{10}\selectfont
       \raggedright {\bfseries Keywords:} #1}
	  }

\renewcommand\section{\@startsection {section}{1}{\z@}%
                   {-3.25ex\@plus -1ex \@minus -.2ex}%
                   {1sp}%
                   {\reset@font\normalsize\bfseries\raggedright}}
\renewcommand\subsection{\@startsection{subsection}{2}{\z@}%
                   {-3.25ex\@plus -1ex \@minus -.2ex}%
                   {1sp}%
                   {\reset@font\normalsize\itshape\raggedright}}
\renewcommand\subsubsection{\@startsection{subsubsection}{3}{\z@}%
                                     {-3.25ex\@plus -1ex \@minus -.2ex}%
                                     {-1em \@plus .2em}%
                                     {\reset@font\normalsize\itshape}}

\newcommand{\ack}[1]{
\section*{Acknowledgments}
\if@anonymous Removed for anonymity \else #1 \fi}

\newcommand{\funding}[1]{
\section*{Funding}
\if@anonymous Removed for anonymity \else #1 \fi}

\newcommand{\coi}[1]{
\section*{Conflict of interest}
\if@anonymous Removed for anonymity \else #1 \fi}

\newcommand{\data}[1]{
\section*{Data availability}
\if@anonymous Removed for anonymity \else #1 \fi}

\newcommand{\roles}[1]{
\section*{Author contributions}
\if@anonymous Removed for anonymity \else #1 \fi}

\newcommand{\suppdata}[1]{
\section*{Appendix}
#1}

\renewcommand{\@makecaption}[2]{\vskip\abovecaptionskip
\sbox\@tempboxa{\fontsize{8}{10}\selectfont {\bfseries #1.} #2}%
\ifdim \wd\@tempboxa >\hsize
\raggedright \fontsize{8}{10}\selectfont {\bfseries #1.} #2\par
\else
\global \@minipagefalse
\hb@xt@\hsize{\hfil\box\@tempboxa\hfil}%
\fi
\vskip\belowcaptionskip}

\let\oldtabular\tabular
\renewcommand{\tabular}{\fontsize{8}{10}\selectfont \oldtabular}

\usepackage{orcidlink} 

\usepackage{harvard} 
\citationmode{abbr} 

\makeatother

\begin{document}

\articletype{Paper} 

\title{Cone-beam artifact reduction in Gamma Knife CBCT images using a line-arc-line scan trajectory}

\author{Alexandra Alain-Beaudoin$^{1,2}$\orcidlink{0009-0001-7599-8448}, Håkan Nordström$^3$\orcidlink{0000-0003-3751-7700}, Luc Beaulieu$^{1,2}$\orcidlink{0000-0003-0429-6366} and Joakim da Silva$^{3,*}$\orcidlink{0000-0002-7771-842X}}

\affil{$^1$Département de physique, de génie physique et d'optique, et Centre de recherche sur le cancer, Université Laval, Québec, Canada}

\affil{$^2$Service de physique médicale et radio-protection, et Axe Oncologie du CRCHU de Québec, CHU de Québec Université Laval, Québec, Canada}

\affil{$^3$Physics and Advanced Applications, Elekta Instrument AB, Stockholm, Sweden}

\affil{$^*$Author to whom any correspondence should be addressed.}

\email{joakim.dasilva@elekta.com}


\input{Document/Abstract}

\input{Document/Intro}
\input{Document/Method}
\input{Document/Resultats}
\input{Document/Discussion}

\input{Document/Conclu}

%
%

\ack{The authors would like to thank Tor Andersson Bakszt, Marianne Plantz and Therese Ullén for their assistance in establishing the experimental setup for this study.}

\funding{This study was supported by Mitacs through the Mitacs Accelerate International program [reference number IT42919]. In this context, Elekta Instrument AB (Stockholm, Sweden) partly funded the study. 
This work was also supported by the Natural Sciences and Engineering Research Council of Canada [funding reference number 596759-2024]; and the Fonds de recherche du Québec [https://doi.org/10.69777/351806].}

\coi{Håkan Nordström and Joakim da Silva are employees of Elekta Instrument AB. Alexandra Alain-Beaudoin receives a stipend partly funded by Elekta Instrument AB. Luc Beaulieu has no conflict of interest to disclose.}


\roles{
Alexandra Alain-Beaudoin: Data curation, formal analysis, funding acquisition, investigation, methodology, software, validation, visualization, writing - original draft, writing - review \& editing
\newline
Håkan Nordström: Conceptualization, resources, writing - review \& editing
\newline
Luc Beaulieu: Funding acquisition, project administration, supervision, writing - review \& editing
\newline
Joakim da Silva: Conceptualization, methodology, project administration, resources, supervision, validation, writing - review \& editing}


\newpage

\suppdata{
\input{Document/Annexe}
}



\input{Document/bbl.tex}
\end{document}

%% file: Document/Abstract.tex
\section*{Abstract}

\textit{Objective.} 
Gamma Knife cone-beam computed tomography (CBCT) images suffer from distinct cone-beam artifacts for some patients, due to the conical X-ray beam which is oriented to intersect the detector perpendicularly at its inferior edge.
The use of an exotic scan trajectory which provides more complete sampling across the field of view than the current arc can reduce cone-beam artifacts. In this study, the feasibility of a line-arc-line scan trajectory for the Gamma Knife CBCT is assessed through a proof of concept. 
\textit{Approach.} 
Using a customized research Gamma Knife, a Catphan 503 and an anthropomorphic head phantom are imaged for different exotic scan trajectories, consisting of combinations of arcs and lines. The CBCT images for each trajectory are then reconstructed with an iterative algorithm.
\textit{Main results.} 
A line-arc-line trajectory provides the largest image quality improvement among the investigated trajectories, with no noticeable cone-beam artifacts in the CBCT images: the axial interfaces between modules of the Catphan are well defined, and the superior edge of the phantom on the axial axis is sharper by 92\%, compared with the current arc trajectory. For the head CBCT images, the proposed trajectory removes most of the cone-beam artifact on the superior side of the skull, characteristic of the current Gamma Knife CBCT images. 
The artifact reduction also leads to a visual improvement in terms of uniformity in both phantoms. 
\textit{Significance.} 
A line-arc-line CBCT scan trajectory would be feasible on the Gamma Knife with limited changes to the current configuration, and could produce images with improved image quality.

%% file: Document/Intro.tex
\section{Introduction}  

Cone-beam computed tomography (CBCT) is a type of medical imaging popular for on-board patient positioning due to its acquisition rapidity, low imaging dose and ease of use \cite{buzug,jaffray_2002_cbct}. However, a compromise for this practicality is poor image quality compared to CT. As the Gamma Knife is used for intra-cranial treatments with stereotactic precision, improving the image quality of its integrated CBCT system continues to be of interest. 

One of the main artifacts in head images from the Gamma Knife CBCT system are cone-beam artifacts on the superior side of the skull, caused by the current circular arc scan trajectory combined with the geometrical configuration of the CBCT system. A particularity of the Gamma Knife CBCT design is the position of the piercing point at the inferior edge of the detector, which creates large cone angles towards the superior side of the detector. A CBCT acquisition made from a single arc aligned with the inferior edge of the detector will therefore produce an image with cone-beam artifacts whose magnitude increases in the superior direction. These geometrical artifacts are object dependent, appearing when high-density variations occur in the axial direction, such as on the superior side of the skull. 
\newpage

While a single arc scan trajectory undersamples the scanned volume, which prevents exact reconstruction inside the FOV, a more complex scan trajectory with better sampling across the FOV may reduce cone-beam artifacts. \citeasnoun{hatamikia_source-detector_2022} published a state-of-the-art review on scan trajectory optimization to achieve various objectives such as reducing cone-beam artifacts, but also extending the FOV, reducing dose and reducing metal artifacts. While tilted orbits can reduce cone-beam artifacts in the axial direction, it does not actually improve sampling, and artifacts may appear in the direction of the rotation axis \cite{wu_cone-beam_2023,zhao_cone-beam_2020}. 
Continuous trajectories made of a combination of lines and ellipses, i.e. helices, can improve sampling, reducing cone-beam artifacts, and extend the FOV \cite{yu_extended_2016,guo_c-arm_2020,yu_line_2011}. 
Recently, saddle trajectories have been proposed to specifically reduce cone-beam artifacts since they allow exact reconstruction \cite{cancelliere_butterfly_2023,hosoo_image_2023,jones_cone-beam_2024,wei_reduction_2024}.
However, they require tilting the arm, which the Gamma Knife CBCT design does not permit;
with the current design, the system's movements are limited to the rotation of the gantry around the z-axis for a fixed 200$\degree$ and the translation of the patient table along the z-axis.

Based on the degrees of freedom of the Gamma Knife CBCT system, theoretical calculations and CBCT simulations were previously performed to assess the potential of various scan trajectories, made up of a combination of arcs, lines and helices \cite{paper1_simulation}. Tomographic incompleteness maps evaluated the level of undersampling at many points in the field of view by measuring the maximal angle between a plane intersecting a given point and the closest source position to that plane. Among the scan trajectories investigated, both approaches indicated that a line-arc-line trajectory is the most promising to reduce cone-beam artifacts. 

The goal of this paper is to demonstrate experimentally the feasibility of the line-arc-line trajectory, and to evaluate the improvements in image quality from measurements on a research Gamma Knife. To this end, an experimental setup is put in place to enable the scan trajectory to be realized on the machine. Measurements are acquired for the line-arc-line trajectory, as well as for other compatible trajectories, using a Catphan 503 (The Phantom Laboratory, Greenwich, United States) and an anthropomorphic head phantom. The CBCT images from the different trajectories are compared in terms of image quality, particularly cone-beam artifacts. After comparison between simulation and experimental results, a conclusion is reached about a proposed scan trajectory for the Gamma Knife CBCT, which is practical for the imaging workflow and which improves image quality.

%% file: Document/Method.tex
\section{Methods}

\subsection{Scan Geometry and Trajectories}

\begin{figure}[htb!]
    \centering
    \includegraphics[width=0.65\linewidth]{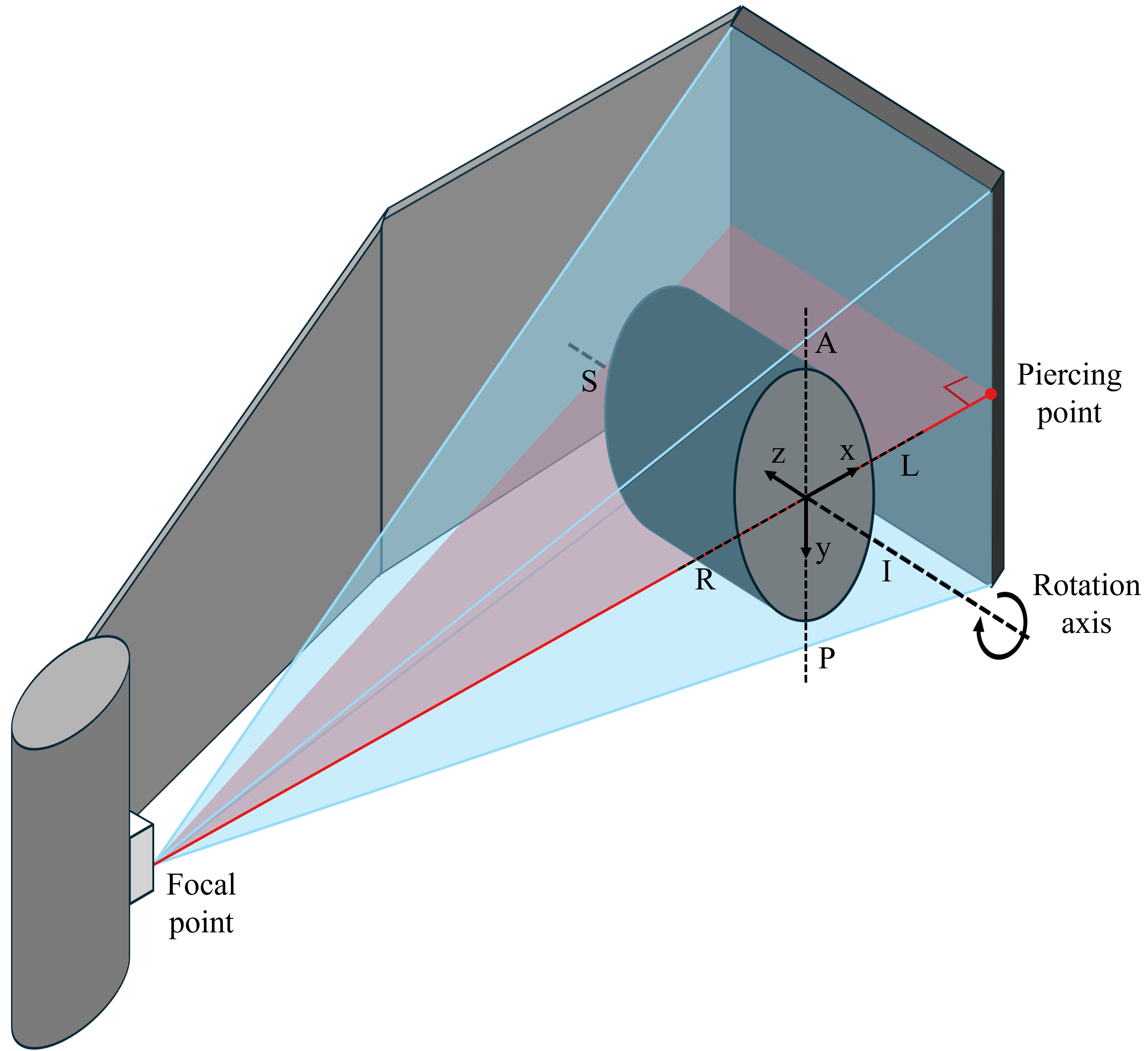}
    \caption{Gamma Knife CBCT geometry, with the imaging coordinate system and the anatomical directions. Reproduced from  \protect\citeasnoun{paper1_simulation}.}
    \label{fig: schema cbct}
\end{figure}

Figure \ref{fig: schema cbct} illustrates the geometry of the Gamma Knife CBCT system for both Icon and Esprit models (Elekta AB, Stockholm, Sweden), with the imaging coordinate system which is fixed in space. The gantry can rotate around the z-axis for an angular range of 200$\degree$. 
The piercing point of the system being at the inferior (I) edge of the detector, the X-ray beam is perpendicular to the detector plane at that point. Towards the superior (S) direction, the beam becomes increasingly inclined compared to the detector, leading to increasingly larger cone-beam artifacts in the reconstruction.

Mechanical constraints of the Gamma Knife design restrict the movements possible for a new scan trajectory. The possible degrees of freedom are limited to translating the patient table and rotating the source over a 200$\degree$ arc around the z-axis. With these movements, the possible scan trajectories are combinations of arcs at various z positions of the table, lines created by moving the table along the z-axis for a certain gantry angle, and helices created by simultaneously translating the table in z and rotating the source. Each segment needs to be efficiently integrated into the trajectory sequence to limit scan time. 
The practicality then prompts to only consider lines formed at the first and last gantry angles of the 200$\degree$ arc, as well as arcs formed at the inferior edge and the middle of the phantom. Also, helical segments are not considered here, because they require simultaneous movement of the table and the arm, making the trajectory more complex; besides, the tomographic incompleteness maps and CBCT simulations have both previously indicated a smaller improvement in image quality compared to other simpler trajectories composed of lines and arcs \cite{paper1_simulation}. 

\newpage

\subsection{Experimental Setup}

\begin{figure}[htb!]
    \centering
    \includegraphics[width=0.8\linewidth]{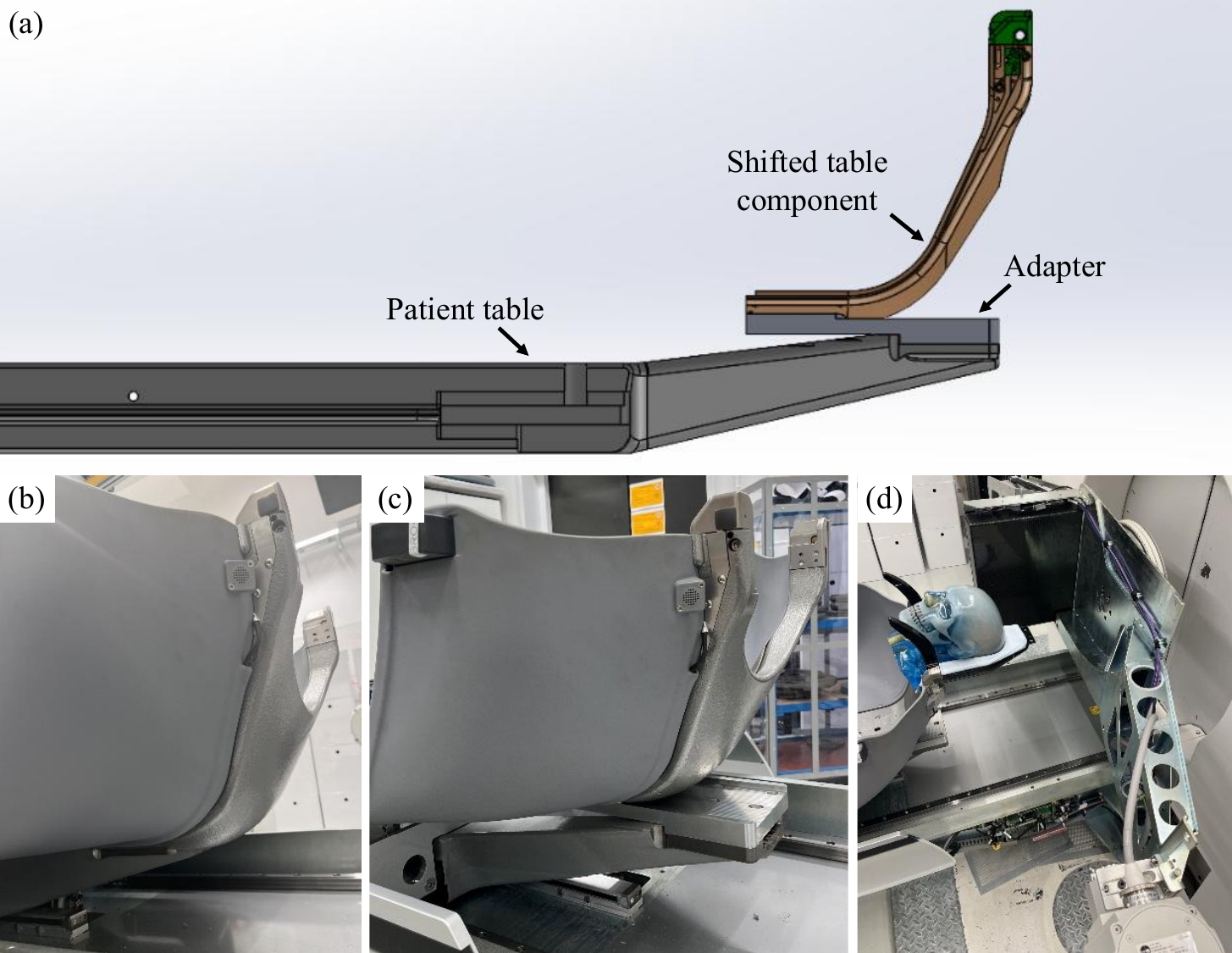}
    \caption{Gamma Knife setup for the investigation of exotic trajectories. (a) shows the digital schema of the setup with the adapter. Photos show the patient table in (b) the standard configuration and in (c) the custom configuration, while (d) shows the patient table with an anthropomorphic head phantom at the most backward position in z in the custom setup.}
    \label{fig: setup}
\end{figure}

The line trajectories require the translation of the patient table along the z-axis from the standard arc position to an inferior position, farther from the radiation unit. However, the current configuration of the Gamma Knife does not permit translating the patient table along the z-axis beyond a few millimetres due to the proximity of the mechanical limit of the range of the patient positioning system. 
A custom setup, shown at Figure \ref{fig: setup}, was put in place to allow the acquisition of projections along a line trajectory: an adapter was designed to shift backwards in z the table piece which fixes the support for the head or phantom. 
The phantom and all obstructing table components moved backwards, the table can translate along a 175-mm range in z, with the most forward position corresponding to the new scan position for the regular arc. The patient table position is increased in y (i.e. downward) to account for the thickness of the adapter.

The patient positioning system is calibrated for the standard scan position. As the custom scan position is different in y and varies in z, a geometrical calibration of the table and the detector is performed with a bearing ball: projections are acquired with the source in anterior, left and right, for multiple table z positions. The measurement of the position of the bearing ball centroid allows characterization of the detector in-plane rotation and translation, as well as the table deviations in x and y when the table is moved along the z-axis. The calibration does not reveal any substantial table deviations in x or y when the patient table is at a different z-position than the standard scan position.

Measurements are done with a Catphan 503 and an anthropomorphic head phantom on a research Gamma Knife Icon (Elekta AB, Stockholm, Sweden), which has the same CBCT hardware and mechanical constraints as the most recent Gamma Knife Esprit (Elekta AB, Stockholm, Sweden). Projections are acquired for 200$\degree$ arcs at the most forward position, which is $z=0$ mm, and at $z=-87.5$ mm, as well as for lines of 175-mm length at the first and last gantry angles with projections every 1.75 mm.

\subsection{CBCT Reconstruction}

Once the projections are acquired, stacks of projections are assembled for different scan trajectories, detailed at table \ref{tab: trajectories}. The CBCT image for each trajectory is then reconstructed with a resolution of 0.5 mm with an iterative algorithm based on weighted least squares (WLS)~\cite{book_wls}, with Total Variation regularization \cite{huber_1964,charbonnier_1997}, developed at Elekta. 
In comparison, the current clinical software employs a reconstruction algorithm for circular scan trajectories of type Feldkamp-Davis-Kress (FDK) \cite{fdk_1984}, which is an approximate filtered backprojection algorithm. The use of the iterative algorithm in this study allows the reconstruction of all scan trajectories with the same algorithm, facilitating comparison between CBCT images from different scan trajectories.

\begin{table}[htb!]
    \centering
    \caption{Description of the scan trajectories investigated, where all scan trajectories except the single arc scan are considered as exotic trajectories.}
    \begin{tabular}{p{2.4cm}|p{12.8cm}}
    \hline
    Scan trajectory & Description \\ \hline
    Arc     &  Standard $200\degree$ single arc at $z=0$ mm with 334 projections\\
    Arc-arc    &  Standard arc followed by a reversed arc with the phantom at $z$~$=$~$-87.5$~mm \\
    Arc-line-arc-line     &  Standard arc at $z=0$ mm, followed by a line at the last gantry angle up to $z=-87.5$ mm, with z-steps of 1.75 mm, followed by a reversed arc with the phantom at $z=-87.5$ mm, followed by a line at the first gantry angle from $z=-87.5$ mm up to $z=-175$ mm, with z-steps of 1.75 mm \\
    Arc-line-arc*-line     & Same as above, but with the reversed second arc made of 33 projections (10\% of a regular arc scan) \\
    Arc-line     &  Standard arc followed by a line at the last gantry angle up to $z$~$=$~$-175$~mm, with z-steps of 1.75~mm \\
    Line-arc-line(1.75)     &  Line at the first gantry angle from $z=-175$ mm up to $z=0$ mm, with $z$-steps of 1.75 mm, followed by a standard arc, followed by a line at the last gantry angle up to $z=-175$ mm, with z-steps of 1.75~mm  \\
    Line-arc-line(3.5)     &  Same as above, except the line segments are made with z-steps of 3.5~mm \\
    Line-arc-line(7)     &  Same as above, except the line segments are made with z-steps of 7~mm    \\
    \hline
    \end{tabular}
    \label{tab: trajectories}
\end{table}

\subsection{Evaluation Metrics}

The magnitude of cone-beam artifacts in the Catphan 503 is assessed through the sharpness of module interfaces and the 20/80 width of the superior edge of the phantom. Specifically, from a 12-mm radius axial area centered on the z-axis, an average profile along the z-axis is computed. As the superior edge of the phantom is the most affected by cone-beam artifacts, they are quantified by means of the width of the superior edge, with limits defined by the positions at 20\% and 80\% of the phantom mean intensity close to the edge. 
Similarly, axial profiles near the 75-mm radial edge of the phantom's inner module are obtained for four opposite regions without material inserts, by averaging the axial planes of cylinders of 18-mm radius. The sharpness in intensity variation between modules and the 20/80-edge width at the superior extremity are evaluated. A schema of the profile locations and their region used for averaging is presented at Figure~\ref{fig: A catphan edge schema} of the Appendix.

%% file: Document/Resultats.tex
\section{Results} 

\begin{figure}[htb!]
    \centering
    \includegraphics[width=1\linewidth]{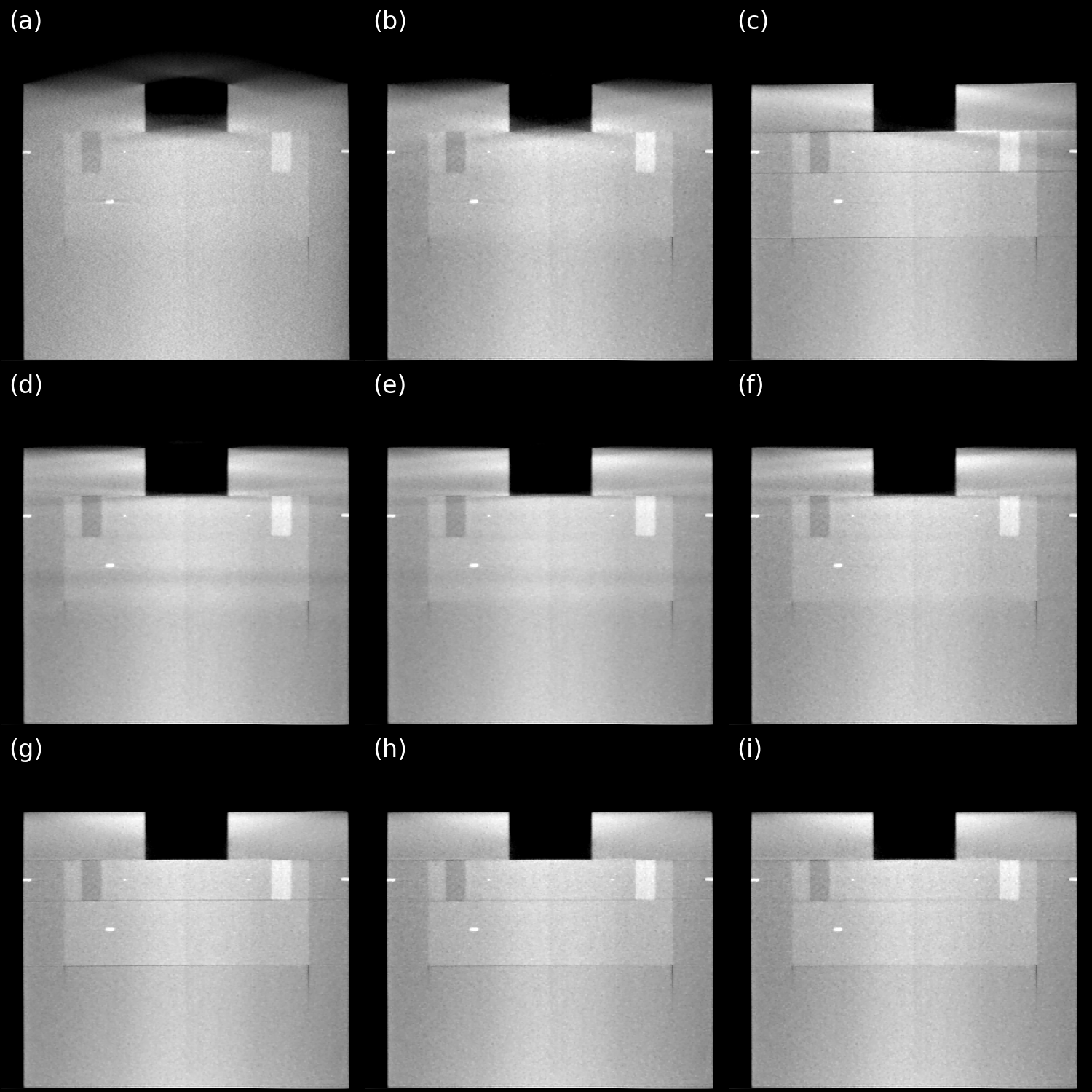}
    \caption{Coronal slices of CBCT reconstructions of the Catphan 503. (a) shows the reconstruction of the arc trajectory with the FDK algorithm. The other reconstructions are with iterative WLS, including weight tapering, for (b) the arc trajectory, (c) the arc-line trajectory, (d) the arc-arc trajectory, (e) the arc-line-arc-line trajectory, (f) the arc-line-arc*-line trajectory, (g) the line-arc-line(1.75) trajectory, (h) the line-arc-line(3.5) trajectory and (i) the line-arc-line(7) trajectory. (window: [-1000,500] HU)}
    \label{fig: A catphan many traj}
\end{figure}

Projection stacks of different scan trajectories are created from the projections acquired on the experimental setup. Figure \ref{fig: A catphan many traj}(a) and (b) show the Catphan reconstructions of the current arc trajectory with the analytic FDK reconstruction and iterative WLS reconstruction, respectively. Both algorithms produce cone-beam artifacts of similar magnitude. In the remainder of this study, only iterative reconstruction is used, to easily compare the reconstructions of different scan trajectories.

\begin{figure}[htb!]
    \centering
    \includegraphics[width=1\linewidth]{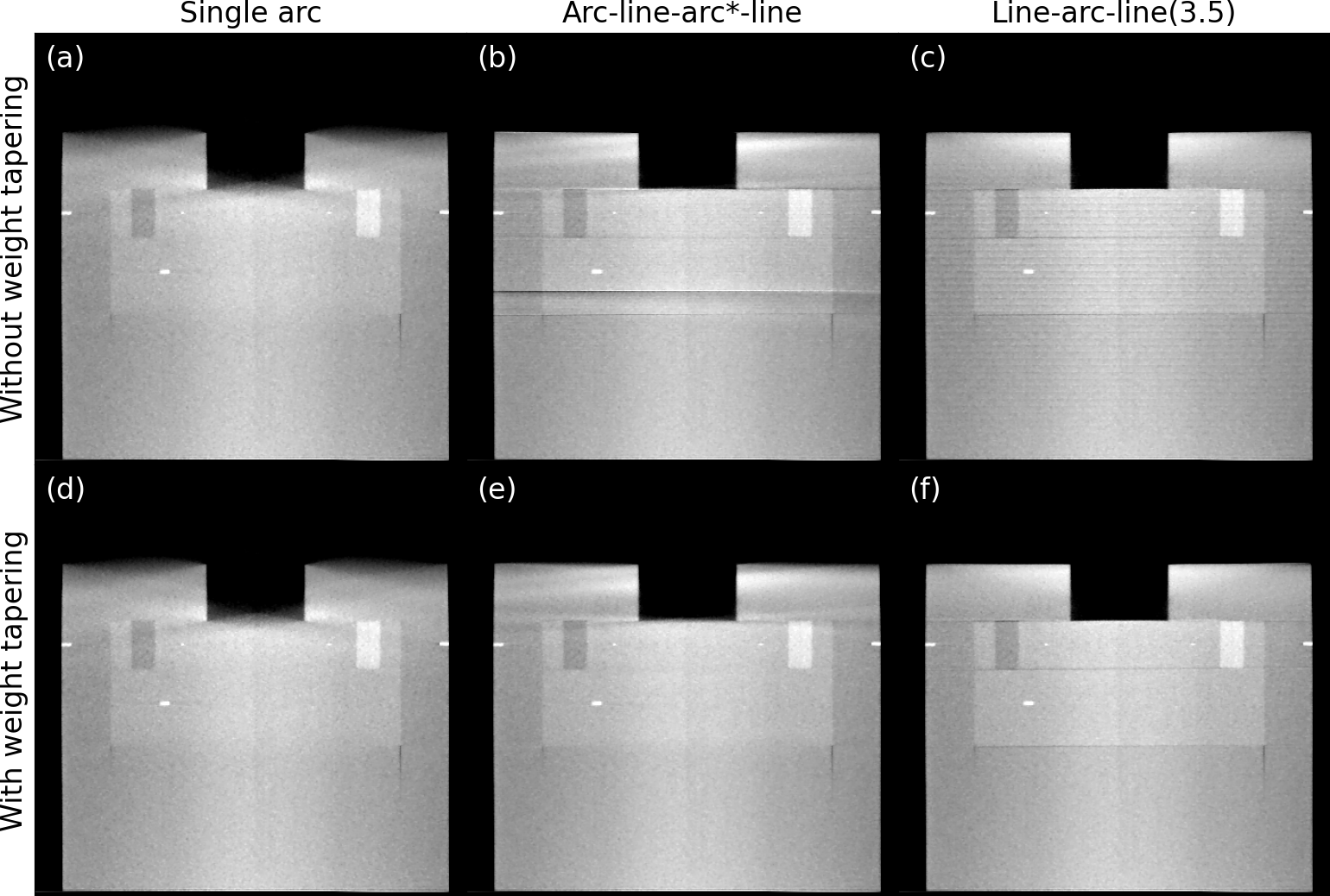}
    \caption{Coronal slices from Catphan 503 CBCT reconstructions for different scan trajectories. In the first row, no weight tapering is applied to the projections during reconstruction, while the second row shows reconstruction with weight tapering applied. (window: [-1000,500] HU)}
    \label{fig: catphan without/with tapering}
\end{figure}

Figure \ref{fig: catphan without/with tapering} shows slices of CBCT reconstructions of a Catphan 503 for the arc, arc-line-arc*-line and line-arc-line(3.5) trajectories. The results from the first row are obtained by applying the original version of the iterative reconstruction algorithm, where the pixels in the projections all have the same weight. 
The reconstructions of the exotic trajectories have new artifacts in the form of a dark band just below the source axial position of the second arc as well as dark streaks at the source position of each projection during the line segments. This new artifact was previously predicted in simulations \cite{paper1_simulation}, and will be referred to as the scatter discrepancy artifact. It is caused by a variation in scatter magnitude when the source is moved axially and the irradiated volume of the object changes. The scatter magnitude decreases as the source moves in the superior direction, and longer source steps will lead to more discrepancy in scatter between the overlapping projections. Consequently, the artifact caused by a second arc is more severe than those caused by line segments. 

Linear weight tapering of the projections was added to the reconstruction algorithm. This tapering is applied to the axially shifted projections and starts from the inferior edge of the projections. While a previous simulation study \cite{paper1_simulation} has shown that a weight tapering of the projections along a range equivalent to 1.2$\times$ the source axial step is sufficient to remove the artifact, experimental results from this work indicate that a longer range of about 1.4$\times$ the source axial step is necessary. As with the simulations, the source axial step used for the range is defined by the axial distance between the arcs if a second are is present, otherwise by the axial distance between each projection of the line segments. The second row of Figure \ref{fig: catphan without/with tapering} shows the reconstructions with the weight tapering applied along the range of 1.4$\times$ the source axial step, where the scatter discrepancy artifact is virtually gone. 

Figure \ref{fig: A catphan many traj} also shows a comparison of reconstructions for all the investigated scan trajectories, where weight tapering was applied during reconstruction of exotic trajectories. Despite the tapering range increase compared to simulations, the reconstructions from trajectories composed of one arc with line segments show virtually no increase of cone-beam artifacts, since the tapering range remains short. However, the wide tapering required for the trajectories with a second arc causes an increase in cone-beam artifacts as predicted.

Overall, by visual inspection of cone-beam artifacts and image uniformity, the trajectories with a second arc (Figure \ref{fig: A catphan many traj} (d,e,f)) have a poorer image quality than trajectories made of line segments before and after the standard arc (Figure \ref{fig: A catphan many traj} (g,h,i)). 
Comparing the arc-arc (d) and arc-line-arc-line (e) trajectories, when a second arc is in the trajectory, adding lines to the trajectory reduces slightly the severity of the remaining dark band artifact. Removing the majority of the projections in the second arc to reach the arc-line-arc*-line (f), the dark band is no longer visible while the remaining cone-beam artifacts are similar. Although the scatter discrepancy artifact is still as severe without weight tapering, the addition of weight tapering virtually removes all of that artifact. 
Comparing the arc-line (c) and line-arc-line(3.5) (h) of same total number of projections, distributing the projections across two near opposite lines improves sampling and helps to achieve a more symmetric image quality. 

The line-arc-line trajectories (g,h,i) have the most improved image quality, where the smaller the source axial step, the sharper the interfaces between the Catphan modules. The artifact reduction in the line-arc-line trajectories also comes with a uniformity improvement in regions of homogeneous material compared to the other exotic trajectories.
These results confirm the previous theoretical analysis through incompleteness maps and CBCT simulations, where a line-arc-line trajectory was the most promising for the Gamma Knife CBCT \cite{paper1_simulation}.

\begin{figure}[htb!]
    \centering
    \includegraphics[width=0.7\linewidth, trim={0 0 0 0.5cm},clip]{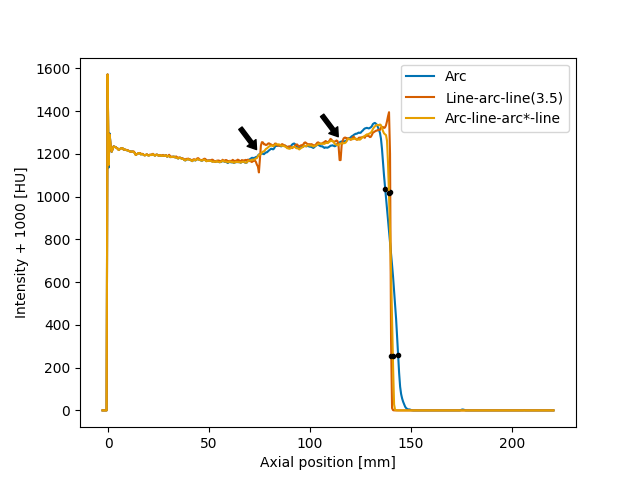}
    \caption{Average profiles of the Catphan 503 near the isocenter axis, for the arc, line-arc-line(3.5) and arc-line-arc*-line trajectories. Black arrows show where there are changes of modules and black points indicate the limits for the measurement of the 20/80 edge width.}
    \label{fig: profile catphan edges}
\end{figure}

To further analyze the image quality improvement of each trajectory, the average axial profiles of the Catphan at the different positions shown in Figure \ref{fig: A catphan edge schema} of the Appendix are obtained. The central profiles of the arc, line-arc-line(3.5) and arc-line-arc*-line trajectories are shown in Figure \ref{fig: profile catphan edges}. 
The interfaces between phantom modules are seen more sharply in the line-arc-line profile than the arc-line-arc*-line and arc profiles, which can be attributed to the absence of cone-beam artifacts. At the superior extremity of the phantom, the edge of the line-arc-line(3.5) is the sharpest, although the arc-line-arc*-line edge is closer to the line-arc-line than the arc trajectory. The 20/80 superior edge widths from more trajectories and for other axial profiles are presented at Table \ref{tab: edge widths}. The widths from the central profile are smaller than those from the other profiles since that edge is positioned lower in axial, resulting in smaller cone-beam artifacts. The limited angular range of the short scan causes a positional dependency in the axial plane, with artifacts more severe towards the anterior direction for a single-arc scan. It is interesting to notice how the edge widths of the arc-line trajectory are more asymmetric than those from the line-arc-line trajectories. There is also a systematic reduction in edge width as the source axial step becomes smaller for the line-arc-line trajectories. 

\begin{table}[htb!]
    \centering
    \caption{Width in mm of the superior edge of the Catphan at different positions in the axial plane, with the ratio to the corresponding width of the arc trajectory in parenthesis. The central edge is at an axial position of 140 mm and the other edges are at 170 mm.}
    \begin{tabular}{c|ccccc}
    \hline
    & \multicolumn{5}{c}{Profile position} \\
    Scan trajectory     &  Central & AR & AL & PL & PR \\
    \hline
     Arc    & 6.37 (1.00) & 9.73 (1.00) & 9.63 
     (1.00) & 9.10 (1.00) & 9.15 (1.00) \\
     Arc-arc    & 2.46 (0.39)  & 4.04 (0.42) & 3.93 (0.41) & 3.75 (0.41)  &   3.91 (0.43)  \\
     Arc-line-arc*-line   & 1.93 (0.30)  & 2.06 (0.21)  & 2.69 (0.28)  & 2.41 (0.27)    &  2.79 (0.30)  \\
     Arc-line    & 0.57  (0.09)  & 1.68 (0.17) &  0.65  (0.067) &  1.79 (0.20)     &  1.90 (0.21)   \\
     Line-arc-line(7)   & 0.73 (0.11)  & 0.95 (0.098)  & 1.03 (0.11) &  1.31 (0.14)    &  1.00 (0.11)  \\
     Line-arc-line(3.5)   &  0.52 (0.081)  &  0.71 (0.073) & 0.70 (0.073) & 1.06 (0.12)  &  0.69 (0.076)   \\
     Line-arc-line(1.75)   & 0.35 (0.056)  &  0.65 (0.067) &  0.65  (0.067)  & 0.98 (0.11)   &  0.45 (0.049)  \\
     
     \hline
    \end{tabular}
    \label{tab: edge widths}
\end{table}

\begin{figure}[htb!]
    \centering
    \includegraphics[width=0.88\linewidth]{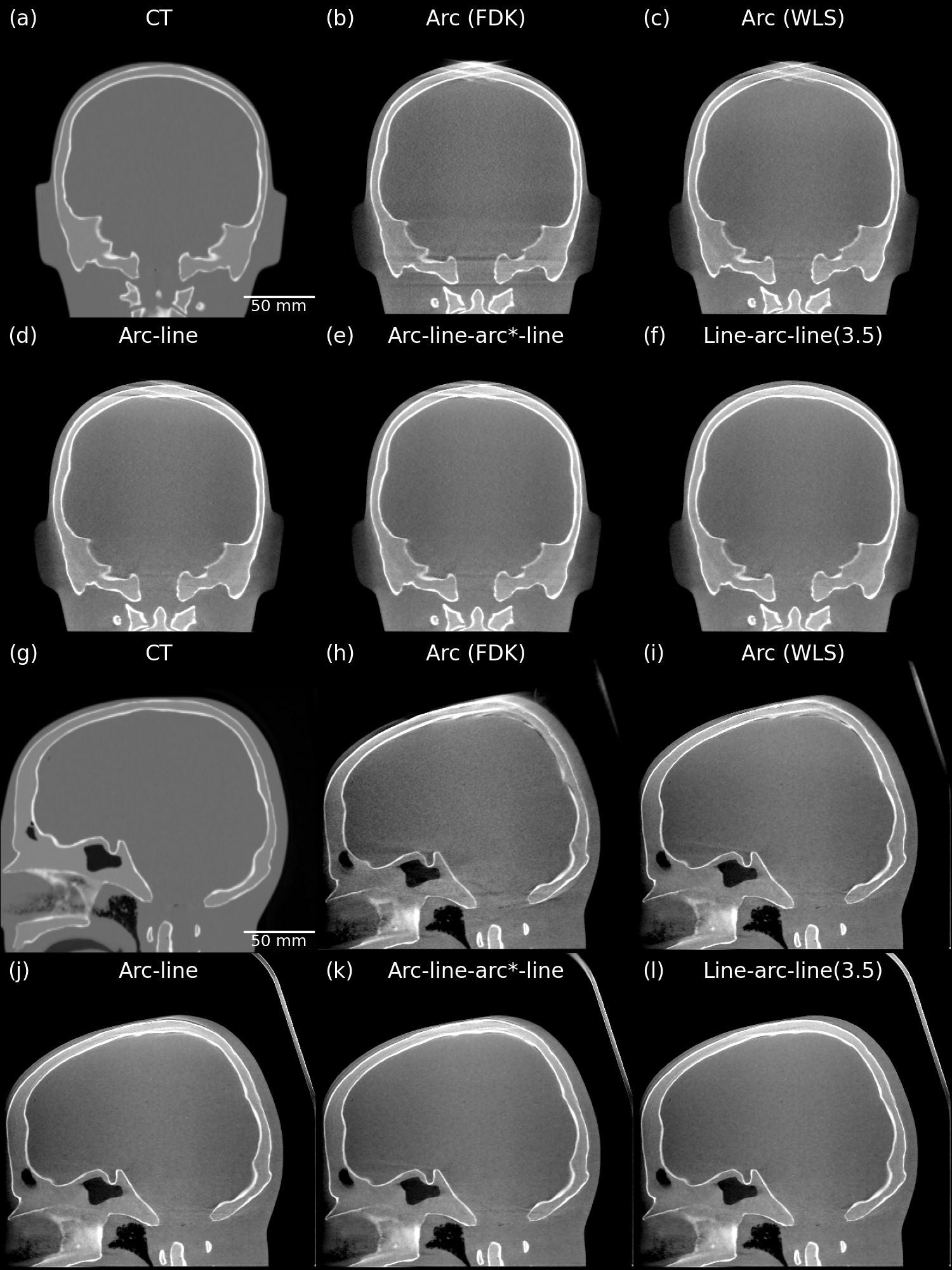}
    \caption{Coronal (a-f) and sagittal (g-l) views of the CT (a,g) and CBCT (b-f,h-l) reconstructions of the anthropomorphic head phantom. (b,h) show an image with FDK reconstruction while (c,i) show the same arc trajectory reconstructed with WLS algorithm. The exotic trajectories shown are the (d,j) arc-line, the (e,k) arc-line-arc*-line and the line-arc-line(3.5). (window: [-1000,1500] HU)}
    \vspace{5mm}
    \label{fig: head results}
\end{figure}

Figure \ref{fig: head results} shows the CBCT reconstructions from the head phantom for selected scan trajectories. The comparison between FDK and WLS reconstructions shows a reduction of dark streaks in the inferior half of the phantom with the iterative reconstruction. Among the exotic trajectories, the line-arc-line shows the smallest cone-beam artifact on the superior side of the skull. Only a thin dark streak remains, possibly due to beam hardening. Uniformity is improved, mostly in soft tissue regions near bone material. 
A small new artifact appears in the sagittal view for all exotic trajectories shown, in the form of streaks in the anterosuperior edge of the phantom which follows the shape of the skull high-density structure. This artifact is attributed to an imperfect geometric calibration of the patient positioning system, which impacts the reconstruction of a CBCT image where the patient table is moved during the scan. 

%% file: Document/Discussion.tex
\section{Discussion} 

A previous theoretical and simulation study \cite{paper1_simulation} has indicated that, out of a variety of scan trajectories compatible with the Gamma Knife, a line-arc-line scan trajectory would reduce cone-beam artifacts the most. In this study, experimental measurements on a customized Gamma Knife confirm the feasibility of the line-arc-line trajectory and that it provides the largest image quality improvement out of the scan trajectories investigated.

Measurements on a Catphan 503 were done for quantitative assessment of the image quality. The CBCT images revealed an artifact caused by scatter discrepancy between axially overlapping projections, which is more severe for overlaps between arcs than those across a line. 
The linear weight tapering applied to the projections during reconstruction removes the scatter discrepancy artifact for the trajectories with additional lines only, and cone-beam artifacts remain virtually absent for the line-arc-line trajectories with a tight source axial step.
To obtain a CBCT image without scatter discrepancy artifacts in trajectories with multiple arcs, while including weight tapering of the projections, the number of projections in the axially shifted projections needs to be reduced, for instance to 10\% of the standard arc's projections as in Figure \ref{fig: catphan without/with tapering}(e). The range along which tapering is applied also has to be increased according to the distance between the arcs, which leads to the re-emergence of some cone-beam artifacts.

\newpage
Analysis of the axial profiles of the Catphan have revealed that a sharp definition of the interface between modules, across the axial plane of the phantom, is visible only with the line-arc-line trajectories. When the width of the superior edge of the Catphan is measured for different positions in the axial plane, there is a systematic reduction in the width as the source step for the line segments becomes shorter. However, this numeric improvement reaches the limit of the human eye. When comparing the Catphan superior edge in the images of the last row of Figure \ref{fig: A catphan many traj}, differences become less distinguishable as the source step becomes shorter. With the aim of limiting scan time and imaging dose, the line-arc-line(3.5), which represents an additional 100 projections to the current trajectory, seems to reach an adequate balance in adding only projections that provide useful information.



For all trajectories, the CBCT images present a hill artifact, where voxel values are higher towards the central axis. This occurs because the bowtie filter of the Gamma Knife CBCT hardens the beam more towards its periphery. The less-hardened beam near the central axis is more attenuated than the peripheral beam of higher average energy, creating a hill artifact in the images. Moreover, the bowtie filter was designed for a cylindrical object with a diameter of an average head. The Catphan being wider than a head, the bowtie filter, which is too narrow, will accentuate the hill artifact. 
These effects from the bowtie filter are then more important than the cupping effect caused by scatter and beam hardening in the object, where, as longer beam paths across the phantom generate more scatter and hardens more the beam, there is an underestimation of the beam attenuation towards the center of the phantom. 

The conclusion regarding the choice of exotic trajectory for the Gamma Knife is upheld when looking at images of a head phantom. The cone-beam artifact on the superior side of the skull, which is characteristic of real head scan images of the Gamma Knife, is virtually removed in Figure~\ref{fig: head results}(f,l) together with the streaks creating non uniformities in soft-tissue on the inferior side of the skull.
Also, while linear weight tapering was not necessary with CBCT simulations of a head phantom \cite{paper1_simulation}, the more severe scatter discrepancy effects that are observed in real CBCT images require application of weight tapering during reconstruction of a head phantom, similarly to the Catphan images.

The current configuration of the Gamma Knife CBCT software does not allow detailed control over exposure parameters. Consequently, all projections were acquired with the same exposure per projection as used in clinical low-dose setting, which corresponds to a CTDI of 2.5~mGy for the arc scan. The reconstructions for different scan trajectories hence have different imaging doses. 
Considering an increase in the number of projections of 30\% with the line-arc-line(3.5) compared to the arc, this represents an increase in imaging dose of less than 30\% since the irradiated volume becomes smaller as the patient table is moved to form the line segments. For a head which has a similar scanned height as the length of the line, it would be expected that the imaging dose increase is around 15\% for the line-arc-line(3.5).


Since image quality is improved with an exotic trajectory such as the line-arc-line, it may be possible to reduce the imaging dose per projection, with the aim of having the same total imaging dose as the current arc trajectory. A small increase in noise is then expected, but the cone-beam artifacts should not be affected, since the severity of the latter is defined by the scan trajectory and the resulting sampling completeness.
In fact, with the noise mainly of quantum and electronic origins for scans of more than 300 projections, \citeasnoun{zhao_noise_2014} have demonstrated that the increase in the number of projections for the same total dose does not impact the quantum noise in the CBCT image, but only increases the electronic noise; with a lower signal per projection, the relative contribution of electronic noise to each projection increases. Based on the measurements by \citeasnoun{zhao_noise_2014}, it is reasonable to expect that, for a voxel size of 0.5 mm and a dose of 2.5 mGy, an increase of 100 projections from an initial scan of 334 projections would increase the noise of a few percents. 

To make the line-arc-line trajectory compatible with the Gamma Knife, the range of the patient positioning system in the direction away from the radiation unit would have to be extended by the desired line length, compared to the current design.
The scanning sequence would start with the patient table at the desired backward z-position, such as 175 mm from the current arc scan position. The table is then moved forward in z, up to the arc scan position, while acquiring projections at the first gantry angle. The standard arc remaining the same as the current arc scan, the scanning sequence ends with the patient table moved backward in z, up to the initial position of the scan, while keeping the gantry at its last angle.

%% file: Document/Conclu.tex
\section{Conclusion}

With the piercing point at the inferior edge of the detector, the use of an arc scan for the Gamma Knife CBCT causes noticeable cone-beam artifacts in the reconstructed images. This study has evaluated experimentally, using a customized Gamma Knife, the feasibility of different exotic scan trajectories consisting of arcs and lines and their potential to improve the CBCT image quality. Among the investigated trajectories, the line-arc-line with a source axial step of 3.5 mm provides the largest image quality improvement in terms of cone-beam artifact reduction and uniformity increase, while minimizing the additional scan time and imaging dose. On a Catphan, the edges along the axial plane, between modules and at the superior extremity, are sharper. On a head phantom, the cone-beam artifact on the superior side of the skull, characteristic of the Gamma Knife CBCT images, is mostly gone. 
From this study's results, one now needs to consider what adjustments would be required to integrate the line-arc-line trajectory into the current imaging workflow, starting with an extended range in z of the patient positioning system, a change of reconstruction algorithm, and a more rigorous geometric calibration of the patient positioning system.

%% file: Document/Annexe.tex
\label{appendix}

\begin{figure}[htb!]
    \centering
    \includegraphics[width=1\linewidth]{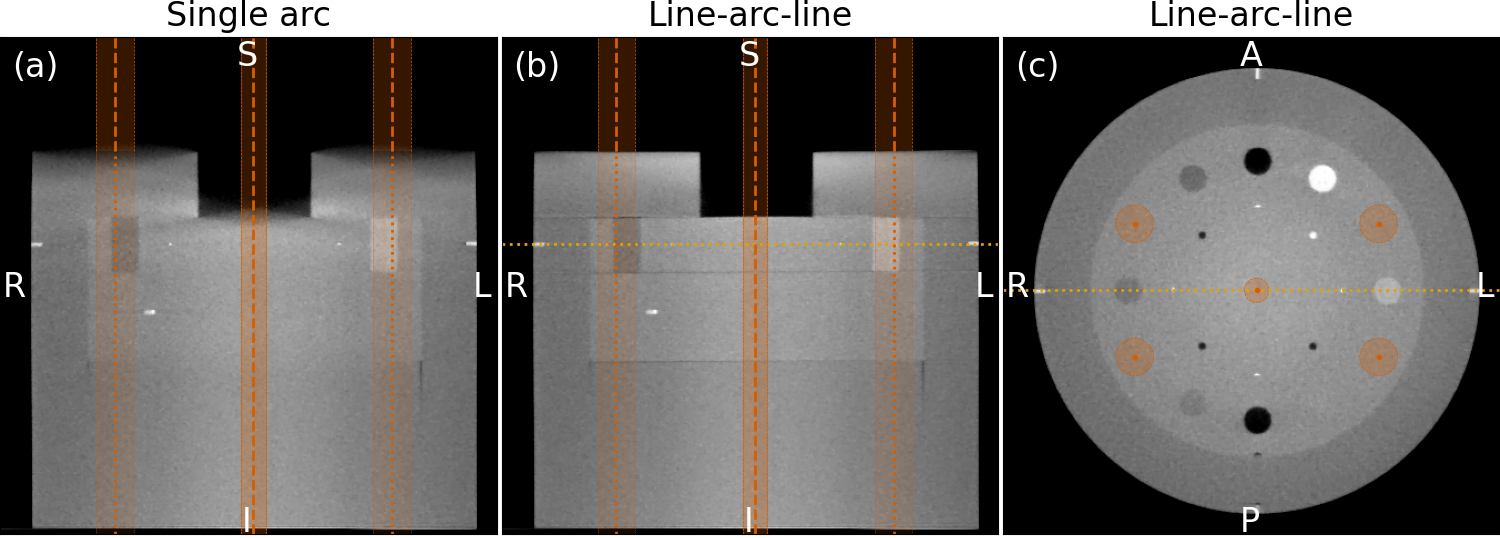}
    \caption{Catphan slices with the regions used to measure average axial profiles, for the arc trajectory (a) and the line-arc-line(3.5) trajectory (b,c). The yellow lines in (b,c) indicate the slice position of the other view.}
    \label{fig: A catphan edge schema}
\end{figure}
